\documentclass[12pt]{iopart}
\usepackage{graphicx}
\usepackage{cite}
\usepackage{iopams}

\begin{document}

\title[{Quasiperiodic oscillations and homoclinic orbits in the NNLS}]{Quasiperiodic oscillations and homoclinic orbits in the nonlinear nonlocal Schr\"odinger equation}

\author{F. Maucher$^1$, E. Siminos$^1$, W. Krolikowski$^2$, S. Skupin$^{1,3}$}

\address{$^1$Max Planck Institute for the Physics of Complex Systems, N\"{o}thnitzer Stra{\ss}e 38, 01187 Dresden, Germany\\
$^2$Laser Physics Centre, Research School of
Physics and Engineering, Australian National University, Canberra, ACT
0200, Australia\\
$^2$Friedrich Schiller University, Institute of Condensed Matter Theory and Optics, 07743 Jena, Germany}

\ead{fabian@pks.mpg.de}

\newcommand{\reffig}[1]{Fig.~\ref{#1}}
\newcommand{\refeq}[1]{Eq.~(\ref{#1})}
\newcommand{\refeqs}[2]{Eqs.~(\ref{#1})-(\ref{#2})}
\newcommand{\ES}[1]{\footnote{ES: #1}}
\renewcommand{\vec}{\mathbf}
\newcommand{\scalarp}[2]{\langle #1,#2\rangle}
\newcommand{\ev}[1]{\hat{e}_{#1}}
\newcommand{\evperp}[1]{e_{#1\perp}}
\renewcommand{\Im}{\mathrm{Im}\,}
\newcommand{\Pperp}[1]{P_{#1\perp}}
\newcommand{\Rot}{\mathcal{R}}

\begin{abstract}
Quasiperiodic oscillations and shape-transformations of higher-order bright solitons in nonlinear nonlocal media have been frequently observed in recent years, 
however, the origin of these phenomena was never completely elucidated. 
In this paper, we perform a linear stability analysis of these higher-order solitons by solving the Bogoliubov-de Gennes equations.
This enables us to understand the emergence of a new oscillatory state as a growing unstable mode of a higher-order soliton. 
Using dynamically important states as a basis, we provide low-dimensional visualizations of the dynamics and identify quasiperiodic and homoclinic orbits,
linking the latter to shape-transformations.
\end{abstract}
\maketitle

\section{Introduction}

Bright solitons are particle-like nonlinear localized 
waves , that keep their form while evolving due to a compensation of diffraction or dispersion of the medium by
 the nonlinear self-induced modification
of the medium~\cite{Agrawal:book:2006}.
Usually, solitons are studied in systems exhibiting local nonlinearities, where the guiding properties of the medium at a particular point in space depend
solely on the wave intensity at that particular point~\cite{Sulem:book:1999}.
Here, we consider nonlocal nonlinearities, i.e. situations in which the nonlinear response of the medium at a point
depends on the wave intensity in  a certain neighborhood of that point, where the extent of this neighborhood is referred to as degree of nonlocality.
Nonlocal nonlinearities are ubiquitous in nature, for example, when the nonlinearity is associated with some sort of transport process, such as
heat conduction in media with thermal response~\cite{Litvak:JETP:1966,Litvak:1975,Davydova:ujp:40:487}, diffusion of charge carriers~\cite{Wright:85,Ultanir:OL:04} or atoms/molecules in
atomic vapors~\cite{Happer:PRL:1977,Suter:PRA:1993}. Nonlinearities are also nonlocal in case of long-range interaction
of atoms in Bose-Einstein condensates (BEC), such as in case of dipolar BEC~\cite{Goral:PRA:05,Tilman:PRL:2005,Beaufils:PRA:2008,Santos:PRL:2005}
or BEC with  Rydberg-mediated~\cite{Henkel:PRL:2010,Maucher:PRL:2011} interactions. In addition, long-range interactions of molecules in nematic liquid crystals also result in  nonlocal
nonlinearities~\cite{McLaughlin:PhysicaD:95,Assanto:IEEEQE:03,Conti:PRL:2003,Peccianti:OL:05}.

The  balance between diffraction and nonlinearity may lead to stable solitons withstanding even  strong perturbations.
In particular, it has been shown, that nonlocal nonlinearities
crucially modify stability properties of localized waves.
With respect to bright solitons, they lead to a much more robust evolution as compared to its local counterpart~\cite{Briedis:OE:05,Lopez-Aguayo:OL:06}.
This is due to the fact, that nonlocality acts like a filter by averaging or smoothing-out  effect on perturbations which would otherwise grow
in case of local response of the medium~\cite{Maucher:PRA:2012}.
For example, higher-dimensional solitons would collapse for systems exhibiting local nonlinearities, whereas they can be
stabilized by nonlocality~\cite{Turitsyn:tmf:85,Bang:pre:2002,Maucher:nonlinearity:2011}.

\begin{figure}
\centerline{\includegraphics[width=0.6\columnwidth]{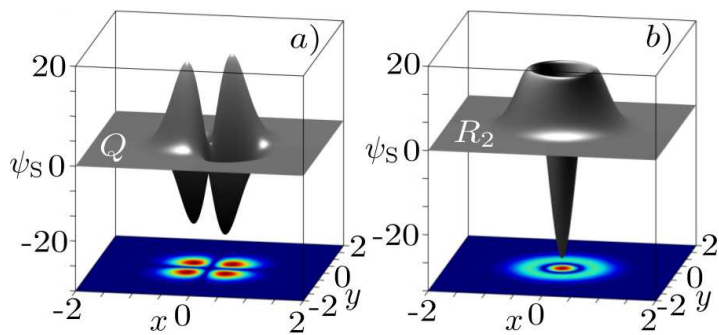}}
\caption{Two particular soliton solutions: a) quadrupole soliton $Q$ and b) second-order radial soliton $R_2$. Both soliton profiles can be chosen real without loss of generality.
The lower plane shows the modulus square depicted in color scale of the two solitons.}
\label{fig:r2_and_q}
\end{figure}

In this work, we investigate the linear stability and nonlinear dynamics of higher-order solitons.
In particular, we study the quadrupole soliton $Q$ and the second-order radial soliton $R_2$ (a hump with a ring), as sketched in~\reffig{fig:r2_and_q}. For those solutions, a quasiperiodic shape transformation between states of different symmetries has been observed recently in~\cite{Bocculiero:PRL:2007,Buccoliero:OE:09}. However,  a complete understanding of this spectacular phenomenon is still missing. One difficulty in the analysis of the shape transformations is that they cannot be described solely in terms of linear perturbation  because  they are not small~\cite{Buccoliero:OE:09}.
Nevertheless, here we show that in spite of the fact that we are dealing with a highly nonlinear phenomenon, deeper understanding can be gained from the linear stability analysis of the corresponding Bogoliubov-de Gennes (BdG) equations. In other words, solutions of the linear stability analysis of the solitons are used  to describe wave  dynamics in the neighborhood of a soliton solution.
Moreover,  in order to fully understand nonlinear dynamics, we employ and further develop techniques recently introduced in dynamical systems studies of dissipative partial differential equations (PDE)~\cite{GHCW07,SCD07}.
These methods employ projection of PDE solutions
from a functional infinite space onto a finite number of important physical states
or dynamically relevant directions.
Here, the relevant directions are mainly the unstable and stable internal modes of the solitons.
The introduction of these low-dimensional projections will allow us to interpret the
non-periodic soliton oscillations 
as indication of homoclinic connections.
Moreover, we are able to understand how different solutions, including quasiperiodic oscillations,
are organized by this homoclinic connection.
The same analysis should also work for a larger variety of higher-order solitons of this nonlocal  system,
such as those presented e.g. in~\cite{Bocculiero:PRL:2007,Buccoliero:OE:09}.

The paper is organized as follows: In Sec.~\ref{secII}, we introduce the governing equations of motion.
In Sec.~\ref{secIII}, we solve the BdG equation to find the internal modes of
the quadrupole soliton $Q$ as well as the second-order radial soliton $R_2$.
In Sec.~\ref{secIV}, we discuss nonlinear soliton propagation, introduce low-dimensional projections and study homoclinic and
quasiperiodic trajectories in this representation. Finally, we will conclude in Sec.~\ref{secV}.

\section{Model equations}\label{secII}

The underlying model equation for our subsequent considerations is the nonlocal nonlinear Sch{\"o}dinger equation (NLS)
\begin{equation}
 i\partial_t\psi + \Delta \psi + \theta \psi = 0,
\label{eq:NLS}
\end{equation}
where $\Delta=\partial_{xx}+\partial_{yy}$ denotes the transverse Laplacian. Depending on the actual context, $|\psi(\vec{r},t)|^2$ can be identified with either the intensity of an optical beam in scalar,
paraxial approximation, or the density of a two-dimensional BEC
within mean field approximation.
The nonlinearity $\theta$ is given by the convolution integral
\begin{equation}
 \theta = \int K(\vec{r}-\vec{r}^\prime) |\psi(\vec{r}^\prime,t)|^2 \mathrm{d}^2 \vec{r}^\prime,
\end{equation}
where the kernel $K$ is determined by the physical system under investigation, and $\vec{r}=(x,y)$.
If $K(\vec{r})=K(|\vec{r}|)$, then \refeq{eq:NLS} is invariant under rotation and the angular momentum is conserved.
This is the case here, as we consider the Gaussian nonlocal model, for which quasiperiodic oscillations have been originally observed~\cite{Bocculiero:PRL:2007,Buccoliero:OE:09}:
\begin{equation}
 K(\vec{r})=e^{-\vec{r}^2}.
\label{eq:gauss}
\end{equation}
Even though there is no actual physical system associated with the Gaussian model, it is commonly used in the literature as a toy model for nonlocal nonlinearities.
Note that without loss of generality the width of the kernel $K$ has been set to unity, in order to have the same scaling as used in~\cite{Bocculiero:PRL:2007,Buccoliero:OE:09}.

\section{Linear stability analysis of higher-order solitons} \label{secIII}

Let $\Phi$ be a bright solitonic solution to our governing equation~(\ref{eq:NLS})
\begin{equation}
 \Phi(\vec{r},t)=\psi_{\mathrm{S}}(\vec{r}) e^{i\lambda t},
 \label{eq:Phi}
\end{equation}
where $\lambda$ is the propagation constant or chemical potential for the case of optical beam or BEC, respectively, and $\psi_{\mathrm{S}}$ denotes the stationary profile of the soliton. Because we will not consider solitons carrying angular momenta (e.g., vortices), we can choose $\psi_{\mathrm{S}}(\vec{r})$ to be real.

In order to find numerically exact stationary profiles $\psi_{\mathrm{S}}(\vec{r})$, we use variational solutions as input to an iterative
solver~\cite{Skupin:PRE:2006}.
Typically, we use a grid of $400\times400$ points to determine $\psi_{\mathrm{S}}(\vec{r})$. This transverse resolution is also employed for numerical integration of \refeq{eq:NLS}, i.e., for beam propagation or time evolution of the two-dimensional BEC.

\begin{figure}
\centerline{\includegraphics[width=0.6\columnwidth]{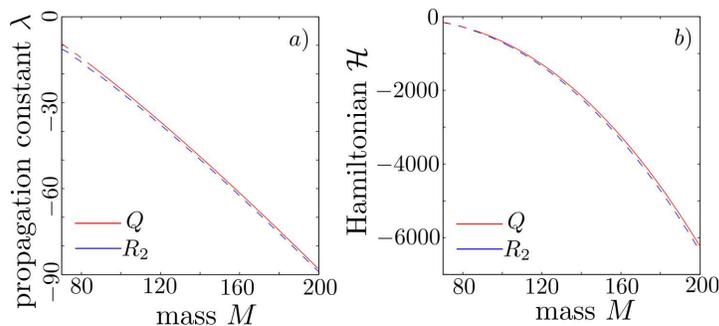}}
\caption{Solitonic family curves for the second-order radial soliton $R_2$ (blue) and
the quadrupole soliton $Q$ (red). Dashed lines indicate parameter domains where the soliton is linearly unstable.}
\label{fig:family_curves_gauss}
\end{figure}

Figure~\ref{fig:family_curves_gauss} shows solitonic family curves or the two higher order solitons we choose to study here, the second-order radial state $R_2$ and the quadrupole $Q$.
Apart from the total angular momentum, there are two conserved functionals,  i.e. the Hamiltonian $\mathcal{H}[\psi]$ associated with invariance with respect to time-translations and
the mass $M[\psi]$ due to a global $U(1)$ phase-invariance:
\begin{eqnarray}
 \mathcal{H}[\psi_\mathrm{S}] & = \int |\nabla \psi_\mathrm{S}|^2\,\mathrm{d}^2\vec{r} - \frac{1}{2}\int |\psi_\mathrm{S}(\vec{r})|^2  K(\vec{r}-\vec{r}^\prime) |\psi_\mathrm{S}(\vec{r}^\prime)|^2 \mathrm{d}^2\vec{r}^\prime \mathrm{d}^2\vec{r},\\
M[\psi_\mathrm{S}] & = \int |\psi_\mathrm{S}|^2 \mathrm{d}^2\vec{r}.
\end{eqnarray}
Obviously, the family curves for the $R_2$ and $Q$ solitons are quite close to each other, which was used in~\cite{Bocculiero:PRL:2007} to explain the observed quasiperiodic shape transformations (energy crossing). However, we will see in the following analysis of projected propagation dynamics in Sec.~\ref{secIV} that this very intuitive picture does not hold.

Let us first recall that the linear stability of solitonic solutions can be studied as an eigenvalue problem as follows.
We introduce a small perturbation $\delta \psi(\vec{r},t)$ to our solitonic solution~$\psi_\mathrm{S}(\vec{r})$ via
\begin{equation}
 \psi(\vec{r},t) = \left[\psi_{\mathrm{S}}(\vec{r})+\delta\psi(\vec{r},t)\right]e^{i\lambda t}
\label{eq:linear_psi}
\end{equation}
Plugging \refeq{eq:linear_psi} into the governing equation~\refeq{eq:NLS} and retaining only first order terms in $\delta \psi$, yields the following
(linear) evolution equation for $\delta\psi$:
\begin{eqnarray}
\left[ i \partial_t -\lambda + \Delta+
 \int K(|\vec{r}-\vec{r}^{\prime}|) \psi_\mathrm{S}^2(\vec{r}^{\prime})\mathrm{d}\vec{r}^{\prime} \right]\delta \psi(\vec{r},t)  \nonumber\\
+ \psi_\mathrm{S}(\vec{r}) \int K(|\vec{r}-\vec{r}^{\prime}|) \psi_\mathrm{S}(\vec{r}^{\prime}) \left[ \delta \psi(\vec{r}^{\prime},t) + \delta
\psi^*(\vec{r}^{\prime},t) \right] \mathrm{d}^2\vec{r}^{\prime} &=0.
\end{eqnarray}
With the ansatz
\begin{equation}\label{eq:ansatz_conj}
 \delta \psi(\vec{r},t) = \delta u(\vec{r})\mathrm{e}^{i\kappa t} + \delta v^*(\vec{r})\mathrm{e}^{-i\kappa^* t}
\end{equation}
for the perturbation we can derive the eigenvalue problem (BdG equation)
\begin{eqnarray}
\label{eq:internal_modes}
 \left[\Delta
 -\lambda + \int K(|\vec{r}-\vec{r}^{\prime}|) \psi_\mathrm{S}^2(\vec{r}^{\prime})\mathrm{d}^2\vec{r}^{\prime} \right] \delta u  (\vec{r})  &  \nonumber\\
 +\psi_\mathrm{S}(\vec{r})\int K(|\vec{r}-\vec{r}^{\prime}|)\psi_\mathrm{S}(\vec{r}^{\prime}) \left[\delta u(\vec{r}^{\prime}) + \delta v(\vec{r}^{\prime}) \right]
  \mathrm{d}^2\vec{r}^{\prime} & = \kappa \delta u(\vec{r})\\
 -\left[\Delta
 -\lambda + \int K(|\vec{r}-\vec{r}^{\prime}|) \psi_\mathrm{S}^2(\vec{r}^{\prime})\mathrm{d}^2\vec{r}^{\prime} \right] \delta v   (\vec{r}) &  \nonumber\\
 -\psi_\mathrm{S}(\vec{r})\int K(|\vec{r}-\vec{r}^{\prime}|)\psi_\mathrm{S}(\vec{r}^{\prime}) \left[\delta v(\vec{r}^{\prime}) +\delta u(\vec{r}^{\prime}) \right]
 \mathrm{d}^2\vec{r}^{\prime} & = \kappa \delta v(\vec{r}).
\end{eqnarray}
Real-valued eigenvalues $\kappa$ of~\refeq{eq:internal_modes} are termed orbitally stable and the corresponding
eigenvector $(\delta u,\delta v)$ can be chosen real-valued. On the other hand, complex eigenvalues with negative imaginary part
indicate exponentially growing instabilities.
We note that due to the special structure of \refeq{eq:internal_modes} [which has its origins in the Hamiltonian structure of \refeq{eq:NLS}], if $\kappa$ is an eigenvalue, then $-\kappa$ as well as $\pm\kappa^*$ are also eigenvalues.

Next, we solve~\refeq{eq:internal_modes} in order to obtain the internal modes of the second-order radial soliton $R_2$ and the quadrupole $Q$, respectively.
A trivial solution to this problem is always given by $(\delta u, \delta v)=\pm (\psi_\mathrm{S},-\psi_\mathrm{S})$ with eigenvalue $\kappa=0$. This so-called trivial phase mode is linked to the phase invariance of solitons. Derivatives of this trivial phase mode with respect to $x$ or \ $y$ are also trivial eigenvectors\footnote{The trivial modes $(\delta u, \delta v)=\pm (\partial_x\psi_\mathrm{S},-\partial_x\psi_\mathrm{S})$ resp.\ $(\delta u, \delta v)=\pm (\partial_y\psi_\mathrm{S},-\partial_y\psi_\mathrm{S})$ are linked to the translational invariance of the system.} with eigenvalue $\kappa=0$, and thus the eigenvalue $\kappa=0$ is degenerate. Moreover, due to symmetry properties of the system trivial modes appear twice in the spectrum, i.e., we expect sixfold degeneracy of the eigenvalue $\kappa=0$. However, when solving the discretized version of~\refeq{eq:internal_modes} numerically, this degeneracy may be lifted. Thus, degenerate eigenvectors with zero eigenvalues may in fact become slightly
complex
without actually indicating an instability. In other words, their nonzero imaginary part is a numerical artefact of the discretization and occurs because the full eigenspace has to be spanned by the eigenvectors.
The actual computation of the linear eigenvalue problem~\refeq{eq:internal_modes} is numerically expensive, since the matrix we have to diagonalize is full, i.e. all entries are nonzero.
In order to achieve reasonable computation times, we usually reduce the grid-size to $100\times100$ points only. Then, the matrix we have to diagonalize has $4 \times 10^8$ non-zero elements.

\begin{figure}
\centerline{\includegraphics[width=0.7\columnwidth]{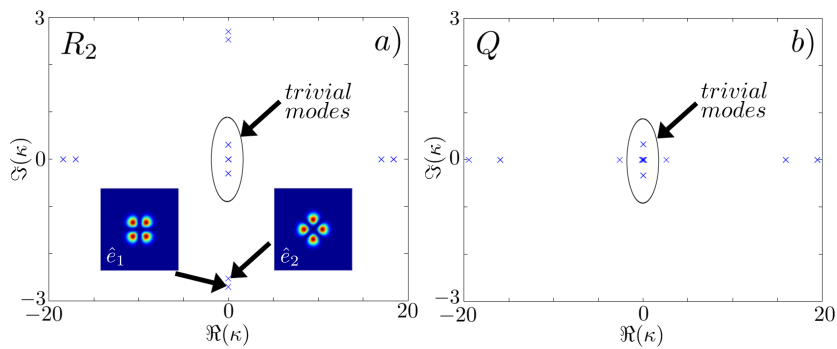}}
\caption{Spectrum of the linear stability analysis (BdG equation) centered around zero for a) the second-order radial soliton $R_2$, and b) the quadrupole $Q$. Both solitons have mass $M=200$.
The radial soliton $R_2$ exhibits instabilities and the unstable eigenmodes $\hat{e}_1(\vec{r})$, $\hat{e}_2(\vec{r})$ resemble quadrupoles [see two insets in a)]; the quadrupole soliton $Q$ is stable.
For both solitons, the degeneracy of the trivial modes is lifted, a numerical artefact due to the discretization of the eigenvalue problem~\refeq{eq:internal_modes}. For sake of readability,
the insets in a) show the absolute square $|\hat{e}(\vec{r})|^2=|\delta u(\vec{r}) + \delta v^*(\vec{r})|^2$ only.}
\label{fig:spectrum_M_200}
\end{figure}

In~\reffig{fig:spectrum_M_200}, we show the spectrum of the linear stability analysis (BdG equation) for the second-order radial soliton $R_2$ and the quadrupole soliton $Q$ [a) resp.\ b)] for mass $M=200$.
Note that for modes with purely imaginary eigenvalue $\kappa=i\Im \kappa$, \refeq{eq:ansatz_conj} reads $\delta \psi(\vec{r},t) = \left[\delta u(\vec{r}) + \delta v^*(\vec{r})\right]\mathrm{e}^{-\Im(\kappa) t}$, and it makes sense to define
\begin{equation}
\hat{e}(\vec{r})=\delta u(\vec{r}) + \delta v^*(\vec{r}).
\end{equation}
Only the second order radial state $R_2$ is unstable, and we name the two unstable internal modes $\hat{e}_1$, $\hat{e}_2$.
The unstable modes $\hat{e}_1$, $\hat{e}_2$ ought to be degenerate for symmetry reasons,
the small splitting of the eigenvalues ($\kappa_1\approx-2.7i$, $\kappa_2\approx-2.5i$) is again a numerical artefact due to the discretization of the eigenvalue problem~\refeq{eq:internal_modes}.
Interestingly, the shape of the unstable eigenmodes $\hat{e}_1(\vec{r})$, $\hat{e}_2(\vec{r})$ resembles quadrupoles.
In fact, for practical purposes (see next section) as well as to verify these findings we furthermore solved the eigenvalue problem~\refeq{eq:internal_modes} for $R_2$ on a radial grid~\cite{Skupin:oe:16:9118}
with eightfold resolution.
Then, instead of two stable and unstable quadrupoles, one finds one stable and unstable vortex with topological charge $m=\pm 2$ and $|\kappa|\approx 2.74$. The vortices corresponding to
$m=2$ and $m=-2$ can be superposed to again yield the quadrupoles $\hat{e}_1$, $\hat{e}_2$ found already with the full 2D solver, but with much higher precision.
Because~\refeq{eq:internal_modes} is linear, the amplitudes of the $\hat{e}_j$ are not fixed, and we normalize the latter according to
\begin{equation}
 \int \hat{e}_j^*(\vec{r}) \hat{e}_j(\vec{r}) \mathrm{d}^2\vec{r} = 1,\hspace{0.5cm} j=1,2.
\end{equation}

\begin{figure}
\centerline{\includegraphics[width=0.3\columnwidth]{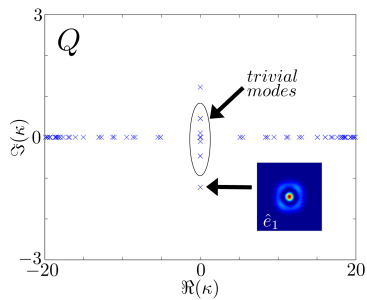}}
\caption{Spectrum of the linear stability analysis (BdG equation) centered around zero for  the quadrupole $Q$ with mass $M=85$. The unstable eigenmode $\hat{e}_1(\vec{r})$ resembles the shape of $R_2$ (see inset), but is of course not rotationally symmetric. Again, the degeneracy of the trivial modes is lifted, a numerical artefact due to the discretization of the eigenvalue problem~\refeq{eq:internal_modes}. For sake of readability, the inset shows the absolute square $|\hat{e}(\vec{r})|^2=|\delta u(\vec{r}) + \delta v^*(\vec{r})|^2$ only.}
\label{fig:spectrum_quadrupole_m_85}
\end{figure}

The quadrupole soliton $Q$ in~\reffig{fig:spectrum_M_200}b) is stable, because all complex eigenvalues correspond to trivial modes and hence the complex form of these eigenvalues is a numerical artefact
as discussed above. However, the quadrupole becomes linearly unstable for $M \lesssim 90$, as indicated in~\reffig{fig:family_curves_gauss} by dashed lines. In \reffig{fig:spectrum_quadrupole_m_85}, we show the results of our numerical stablility analysis for the   quadrupole soliton $Q$ with mass $M=85$. Interestingly, the unstable mode $\hat{e}_1$ with $\kappa_1\approx-1.2i$ resembles the second-order radial soliton $R_2$, i.e., a hump with a (modulated, i.e. not rotationally symmetric) ring.

\section{Projected nonlinear dynamics}  \label{secIV}

\begin{figure}
\centerline{\includegraphics[width=\columnwidth]{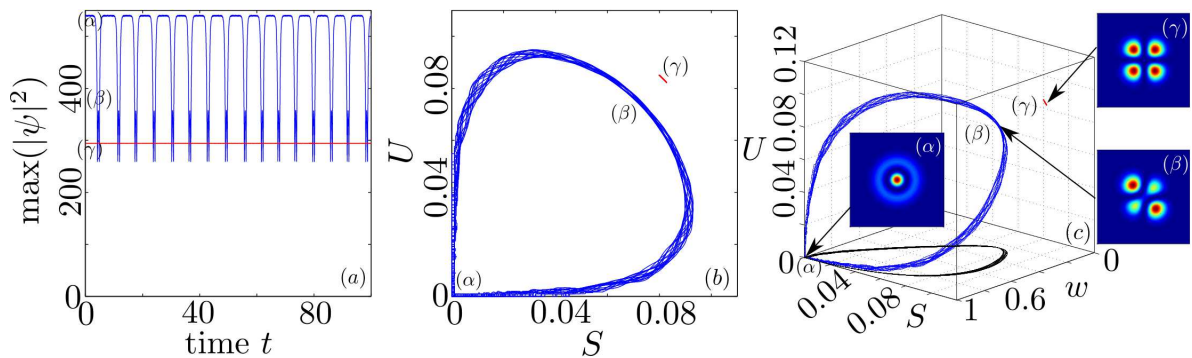}}
\caption{a) Evolution of the peak-intensity of the second-order radial soliton $R_2$ with mass $ M=200$ (upper blue curve).
As expected from the stability analysis~\reffig{fig:spectrum_M_200}~b), the peak-intensity of the quadrupole soliton $Q$ with same mass (lower red curve) is constant during propagation.
Figure b) shows the projected dynamics in the variables $U(t),S(t)$ [see \refeq{eq:US}] for
initial conditions $R_2$ [blue curve, starting at ($\alpha$)] and $Q$ [red curve, starting at ($\gamma$)].
The shape of the former curve hints at a \emph{homoclinic connection}, where the homoclinic point corresponds to $R_2$ ($\alpha$).
Figure c) presents the same dynamics as b), with an additional dimension given by the variable $w$ [see \refeq{eq:projr2}]. 
In this three-dimensional projection, the distance between the quadrupole $Q$ ($\gamma$) and the ''turning point'' ($\beta$) becomes apparent. 
For reasons of clarity, the 3D-dynamics (blue) is additionally projected into $(S,w)$-plane (black), and the orbit of the the quadrupole is 
again shown in red.
The three insets show snapshots of the nonlinear dynamics. 
}
\label{fig:2d_s_over_u}
\end{figure}

The typical dynamics for $R_2$ (here for $M=200$) as an initial condition is shown in \reffig{fig:2d_s_over_u}~a).
To determine the shape of $R_2$, we use the iterative solver mentioned above on a grid containing $400\times400$ points, and we use the same grid for the actual propagation.
As we have seen in Sec.~\ref{secIII}, the second-order radial soliton $R_2$ is unstable over the whole range of mass $M$ and therefore any perturbation,
that has a non-zero overlap with the unstable internal modes $\hat{e}_1$, $\hat{e}_2$ will lead to an
exponential growth of the latter.
Practically, the residual in numerical determination of $R_2$ as well as the propagation algorithm
based on the Fourier split-step method~\cite{Agrawal:book:2006} lead to inevitable numerical noise when propagating and  therefore trigger the instability without adding any additional perturbation.
In our case, however, we added the eigenmode $\hat{e}_1$ as initial perturbation with tiny amplitude $\sim10^{-4}$ to the soliton $R_2$ to control the breakup in a preferred direction.
For small times the dynamics is governed by the exponential growth of the unstable internal mode $\hat{e}_1$, while for later times the evolution becomes highly non-linear,
exhibiting oscillations between $R_2$ [see inset ($\alpha$) in \reffig{fig:2d_s_over_u}~a)] and a state that resembles the quadrupole soliton $Q$ [see inset ($\beta$) in \reffig{fig:2d_s_over_u}~a)]~\cite{Bocculiero:PRL:2007}.
This state $(\beta)$ we will call the ''turning point''. In the following we will examine in detail
the origin and properties of these oscillations.

\subsection{Projection methods}
\begin{figure}
\centerline{\includegraphics[width=0.7\columnwidth]{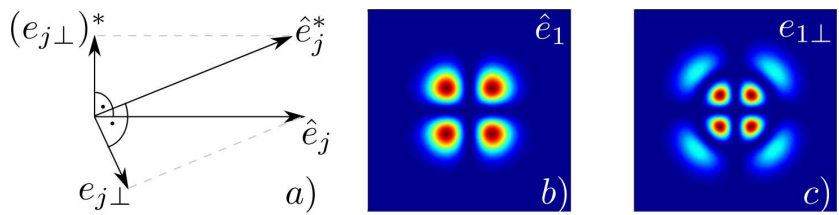}}
\caption{Schematic sketch of the relation between ${\ev{j},\ev{j}^*},\evperp{j}$, and $(\evperp{j})^*$.
By construction, $\evperp{j}$ is orthogonal to the stable eigenvector $\ev{j}^*$, and $(\evperp{j})^*$ is orthogonal to the unstable eigenvector $\ev{j}$.
It is worth to notice that $\evperp{j}$ and $(\evperp{j})^*$ are not orthogonal to each other.
b) and c) show the modulus squared of the internal mode $\hat{e}_1$ and $e_{1\perp}$, respectively.}
\label{fig_scheme_ortho}
\end{figure}

Let us now introduce the projection method mentioned in the introduction~\cite{GHCW07,SCD07} and adopt it to our problem.
To this end, we recall the scalar product of two complex functions $f$ and $g$, defined as
\begin{equation}
 \scalarp{f}{g}  = \int f^*(\vec{r}) g(\vec{r})\mathrm{d}^2 \vec{r}.
\label{eq:sk_prod}
\end{equation}
Obviously, the (unstable) internal modes $\ev{j}$ of $R_2$ introduced before [see~\reffig{fig:spectrum_M_200}] are not orthogonal to
their complex conjugate (stable) $\ev{j}^*$ ($j=1,2$) counterparts
with respect to this inner product.
In other words, stable and unstable eigenspaces $E^{s}$ and $E^{u}$
spanned by eigenfunctions $\{\ev{1}^*,\,\ev{2}^*\}$ resp.\ $\{\ev{1},\,\ev{2}\}$ are not mutually orthogonal.
Thus, straightforward projections onto $\ev{j}$ and $\ev{j}^*$ do not help to elucidate
the propagation dynamics.
To overcome this difficulty we introduce a set of functions which is biorthogonal to ${\ev{j},\ev{j}^*}$ using a Gram-Schmidt-like technique as follows.
First, we define
\begin{equation}\label{eq:evperp}
\evperp{j}=\ev{j}-\scalarp{\ev{j}^*}{\ev{j}}\ev{j}^*,
\end{equation}
which is simply the projection of the unstable
eigenmode $\ev{j}$ onto the orthogonal complement of the stable eigenmode $\ev{j}^*$.
Second, we note that $(\evperp{j})^*=\ev{j}^*-\scalarp{\ev{j}}{\ev{j}^*}\ev{j}$ corresponds to projection of the stable eigenmode $\ev{j}^*$ onto the orthogonal complement of the unstable eigenmode $\ev{j}$.
Then, it is easy to verify biorthogonality of $\evperp{j},(\evperp{j})^*$ with respect to ${\ev{j},\ev{j}^*}$:
\begin{eqnarray}
 \scalarp{\ev{j}}{(\evperp{j})^*} &= 0
 \scalarp{\ev{j}}{\evperp{j}} &\neq 0
\label{eq:basis}
\end{eqnarray}
In~\reffig{fig_scheme_ortho}a) a schematic sketch of the relation between ${\ev{j},\ev{j}^*},\evperp{j}$, and $(\evperp{j})^*$ is depicted,
and~\reffig{fig_scheme_ortho}b-c) show $\hat{e}_1$ and $e_{q\perp}$ explicitly.
It is worth to notice that $\scalarp{\evperp{j}}{(\evperp{j})^*} \neq 0$, i.e., $\evperp{j}$ and $(\evperp{j})^*$ are not orthogonal to each other.

In order to analyze the propagation dynamics of a
solution $\psi(x,t)$ of \refeq{eq:NLS}, we introduce the quantities
\begin{equation}
  U_j = \scalarp{\evperp{j}}{\psi}, \qquad S_j = \scalarp{(\evperp{j})^*}{\psi}.
\end{equation}
By construction, $U_j$ is associated with the unstable eigenmode only ($\evperp{j}$ is orthogonal to the stable one), while $S_j$ is associated with the stable eigenmode only.
Finally, for $R_2$, the two unstable eigenvectors $\ev{1}$, $\ev{2}$ are degenerate (due to rotational symmetry about the origin),
therefore we introduce the rotationally invariant projected variables
\begin{equation}\label{eq:US}
  U(t) = \sqrt{\sum_{j=1}^2 |U_j|^2}\,,  \qquad S(t) = \sqrt{\sum_{j=1}^2 |S_j|^2}\,.
\end{equation}
Then, any pair of wavefunctions $\psi_1(x,t)$ and $\psi_2(x,t)$ related through a rotation amounts to the same value of $U(t)$ and $S(t)$. For a rigorous proof, see Appendix~\ref{appeA}.

\subsection{Indication of homoclinic connections} \label{sec:homo}

Figure~\ref{fig:2d_s_over_u}~b) illustrates the dynamics shown in \reffig{fig:2d_s_over_u}~a) in the variables $S(t),U(t)$ introduced in \refeq{eq:US}.
We clearly see the second-order radial soliton $R_2$ ($\alpha$) decaying into
a quadrupole-like state ($\beta$), the ''turning point'', and then coming back to $R_2$. In the vicinity of $R_2$, the decay starts via the local unstable eigenspace $E^u$ (i.e., $U(t)>0$, $S(t)\approx0$), and the revival of $R_2$ happens via the local stable eigenspace $E^s$ (i.e., $S(t)>0$, $U(t)\approx0$).
The fact that the system repeatedly returns (close) to its initial state $R_2$ and remains at this point some finite, non-constant time with (nearly) zero velocity, hints at the existence of a \emph{homoclinic connection}. A homoclinic connection is a solution which is asymptotic to $R_2$ both in the
$t\rightarrow\infty$ and $t\rightarrow-\infty$ limit.
The time-span, in which the solution remains close to its initial state $R_2$, i.e. the homoclinic point ($\alpha$) in~\reffig{fig:2d_s_over_u}~b),
with practically zero velocity, corresponds to intervals with maximum (nearly) constant peak-intensities in~\reffig{fig:2d_s_over_u}~a).
Because we added a small perturbation in the direction of the eigenmode $\hat{e}_1$ to the initial condition $R_2$, and the presence of numerical noise in general,
we do not see the exact homoclinic connection in our numerical simulations; as the trajectory
comes back towards $R_2$ along $E^s$, there is always a small perturbation along the unstable eigenspace $E^u$ and the trajectory leaves the neighborhood of
$R_2$ to return to it later on. We want to stress here that the existence of homoclinic connections is by no means anticipated in general; our numerical results however indicate the existence
of such homoclinic connections and their persistence along a large range of the mass $M$.

To further illustrate that the ''turning point'' ($\beta$) is indeed well-separated from the quadrupole soliton $Q$ ($\gamma$), we introduce a third variable $w$ by projecting the solitonic wave function $\psi$ onto the radial soliton $R_2$,
\begin{equation}
 w(t)=\frac{\left|\scalarp{R_2}{\psi}\right|}{\scalarp{R_2}{R_2}}. \label{eq:projr2}
\end{equation}
Obviously, for $\psi=R_2$ we find $w=1$,
while for  $\psi=Q$ for symmetry reasons we have $w=0$.
Figure~\ref{fig:2d_s_over_u}~c) shows the resulting projected dynamics on the variables $U,S,w$. We clearly recognize similarities with \reffig{fig:2d_s_over_u}~b),
however, it becomes much more clear how the solution evolves from its origin ($\alpha$) and becomes much more ``quadrupole-like'' in ($\beta$).
In particular, the important separation between the quadrupole-like ''turning-point'' ($\beta$), which still maintains a nonzero projection on $R_2$ and the quadrupole soliton $Q$ ($\gamma$) becomes evident.

\subsection{Quasiperiodic motion} \label{sec:quasi}

In the previous section, we argued that due to numerical limitations, we cannot actually track the homoclinic orbit precisely, but what we find are trajectories that are
very close to the homoclinic connection.
In the present section we will further probe the dynamical importance of the homoclinic orbit by studying trajectories adjacent to it.
In a sense, the ''turning point'' ($\beta$) of the homoclinic orbit is a
state ''in between'' $R_2$ and $Q$. Here we will investigate the dynamics of such ''in between'' states obtained by perturbing
the homoclinic orbit at the ''turning point'' ($\beta$). The perturbations we will consider are not necessarily small and, as we will see, they typically lead
to quasiperiodic oscillations.

A homoclinic orbit is obtained by (slightly) perturbing the initial wavefunction of $R_2$ in the direction of one of the unstable modes (e.g. of $\ev{1}$)
and integrating \refeq{eq:NLS} forward in time. Choosing the direction of the initial perturbation fixes the ``orientation'' of
the subsequent dynamics, and we can thus
decompose the wavefunction at the turning point [point ($\beta$) in~\reffig{fig:2d_s_over_u}]
$t_\mathrm{t}$ into a part parallel to the quadrupole soliton $Q$
and a remainder $L$
\begin{equation}
 \psi(\vec{r},t_\mathrm{t}) = c_\mathrm{Q} Q(\vec{r}) + L(\vec{r}),
\label{eq:decomposition}
\end{equation}
where $c_\mathrm{Q}=\scalarp{Q}{\psi(\vec{r},t_\mathrm{t})}/\langle Q,Q\rangle$ was introduced\footnote{More generally, if the direction of the breakup is arbitrary, one may generalize~\refeq{eq:decomposition} by
decomposing $\psi(t_\mathrm{t})$ into two quadrupoles $Q_1,{\ }Q_2$, where $Q_1$ is rotated by $\pi/4$ with respect to $Q_2$, via
$\psi(\vec{r},t_\mathrm{t}) = c_{\mathrm{Q}_1} Q_1(\vec{r}) + c_{\mathrm{Q}_2} Q_2(\vec{r}) + L(\vec{r})$.}.
Perturbed wavefunctions $\psi_\Gamma(\vec{r})$ are then constructed through
\begin{eqnarray}
\label{eq:psi_alpha}
 \psi_\Gamma^\prime(\vec{r}) &= c_\mathrm{Q} Q(\vec{r}) + \Gamma L(\vec{r}) \\
 \psi_\Gamma &= \sqrt{\frac{\langle \psi,\psi\rangle}{\langle \psi_\Gamma^\prime,\psi_\Gamma^\prime\rangle}} \psi_\Gamma^\prime
 \label{eq:psi_alpha2}
\end{eqnarray}
where $\Gamma$ parametrizes mixed states between $R_2$ and $Q$,
and, in \refeq{eq:psi_alpha2}, the wavefunction was normalized.
Clearly, for $\Gamma=1$, the homoclinic trajectory of $R_2$ can be recovered, whereas of $\Gamma=0$,
the quadrupole soliton is recovered.
In the following the time evolution of the function $\psi_\Gamma$ will be studied.

\begin{figure}
\centerline{\includegraphics[width=0.66\columnwidth]{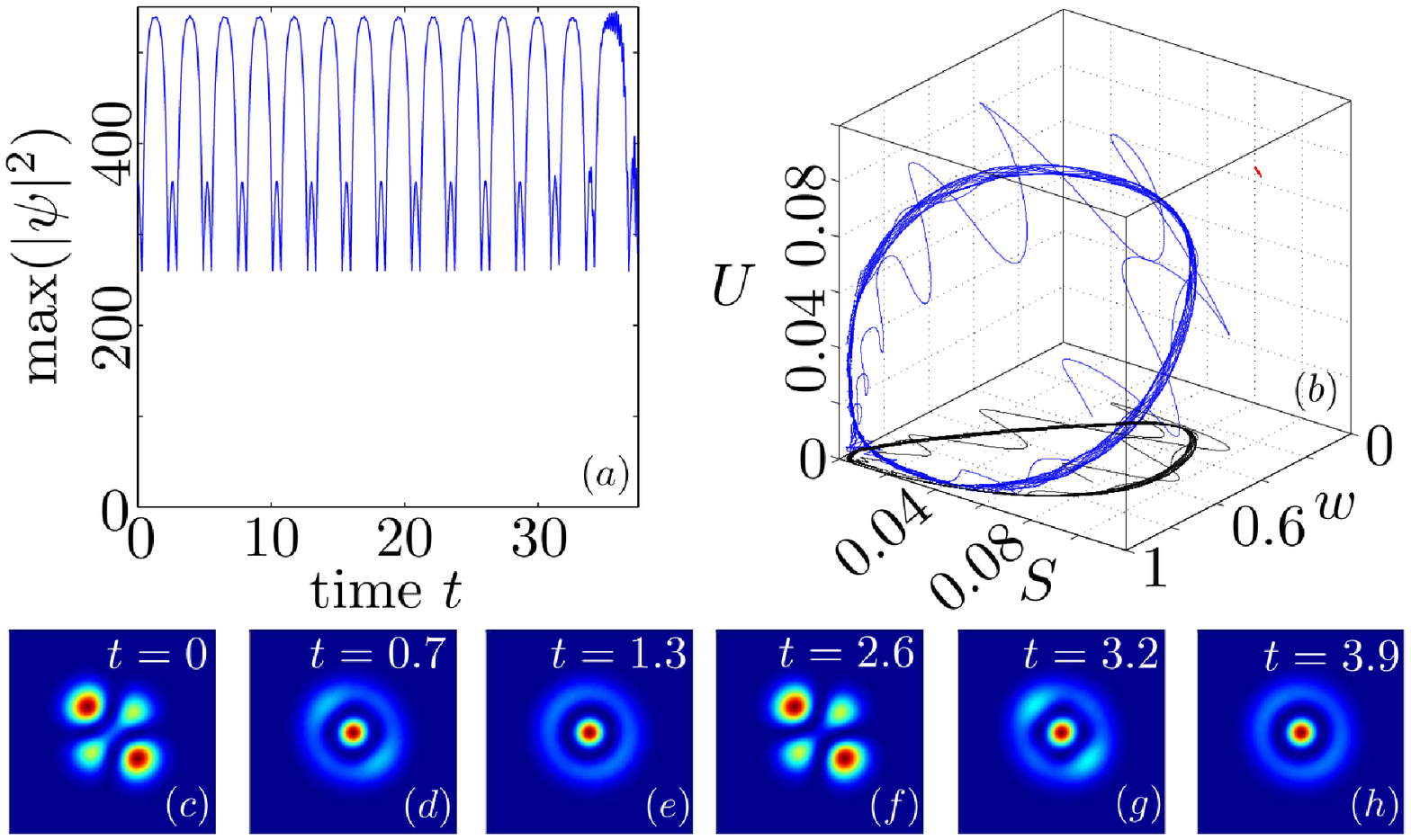}}
\caption{Evolution of $\psi_\Gamma$ for $\Gamma=1.01$ defined in~\refeq{eq:psi_alpha}.
(a) shows the peak-intensity, (b) the orbit in lower-dimensional $S,U,w$ representation, and (c-h) snapshots of the dynamics.
The coloring is the same as in~\reffig{fig:2d_s_over_u}, where the blue curve again represents the actual 3D dynamics and the black 
curve its projection on the $(s,w)$-plane, and the red curve is the orbit of the quadrupole.}
\label{fig:101}
\end{figure}

Let us first consider the dynamics for $\Gamma=1.01$ as shown in~\reffig{fig:101}, which indicates
quasiperiodic behavior for small times (up to $t\simeq25$).
The time spend by this orbit close to $R_2$ is much smaller than for the homoclinic connection, of the previous section.
This becomes apparent when comparing the peak-intensity evolution in \reffig{fig:101}a) with the one in \reffig{fig:2d_s_over_u}a).
In the $(U,S,w)$ projection, this fact results in a smoother curve close to the origin (whereas for a homoclinic
connection a kink appears as $R_2$ is approached, while the ``velocity'' approaches zero).
On the other hand, in the intensity representation, \reffig{fig:101}c-h), the difference between homoclinic and quasiperiodic behavior is
much harder to discern.
Propagation in \reffig{fig:101}a) and b) is shown until $t=35$, when the dynamics already deviates from the quasiperiodic orbit,
indicating that the latter is unstable. This behavior hints to the existence of some chaotic region in state-space,
an issue that will be studied elsewhere.

\begin{figure}
\centerline{\includegraphics[width=0.66\columnwidth]{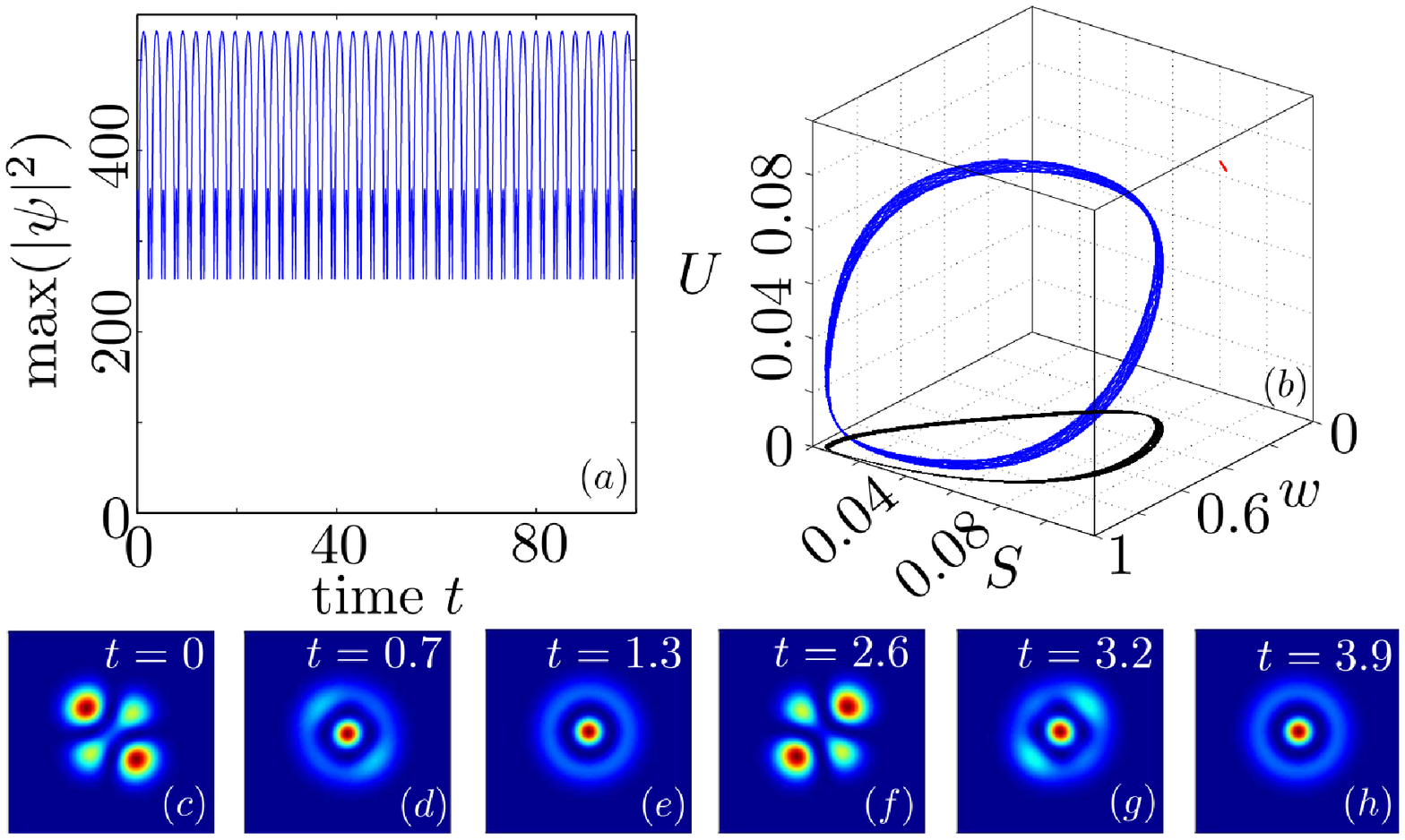}}
\caption{Evolution of $\psi_\Gamma$ for $\Gamma=0.99$ defined in~\refeq{eq:psi_alpha}.
(a) shows the peak-intensity, (b) the orbit in lower-dimensional $S,U,w$ representation and (c-h) snapshots of the dynamics.
The coloring is the same as in~\reffig{fig:2d_s_over_u}, where the blue curve again represents the actual 3D dynamics and the black 
curve its projection on the $(s,w)$-plane, and the red curve is the orbit of the quadrupole.}
\label{fig:099}
\end{figure}

On the other hand, the dynamics for $\Gamma=0.99$, shown in~\reffig{fig:099}, appear again quasiperiodic (see also the
discussion of the Fourier spectra in Sec.~\ref{fourier}), but in this case the orbit appears stable, as it persists
at least up to $t=1500$. The qualitatively different behaviour for $\Gamma=1.01$ and $\Gamma=0.99$ with respect to stability further corroborates the importance of the homoclinic solution $\Gamma=1.00$ ($R_2$). In a certain sense, the homoclinic orbit ''organizes'' regions of stability in parameter space. However, the homoclinic orbit should not be seen as a kind of ''boundary'' between regions of different stability behaviours, because it is just a one-dimensional line in the highly-dimensional parameter space.

\begin{figure}
\centerline{\includegraphics[width=0.66\columnwidth]{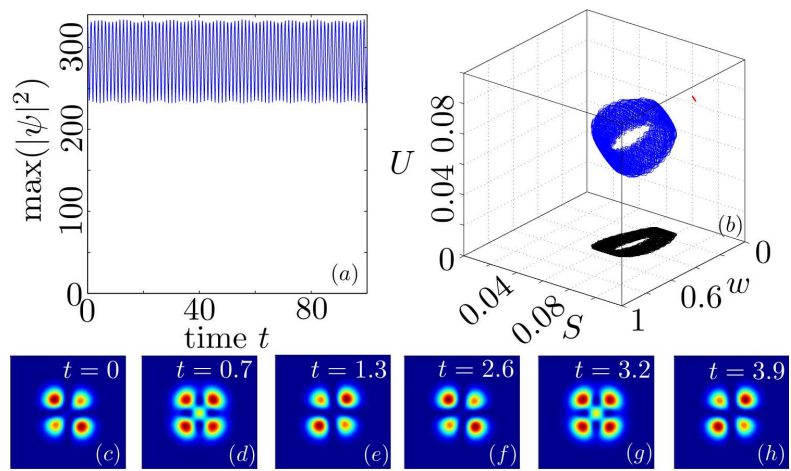}}
\caption{Evolution of $\psi_\Gamma$ for $\Gamma=0.5$ defined in~\refeq{eq:psi_alpha}.
(a) shows the peak-intensity, (b) the orbit in lower-dimensional $S,U,w$ representation and (c-h) snapshots of the dynamics.
The coloring is the same as in~\reffig{fig:2d_s_over_u}, where the blue curve again represents the actual 3D dynamics and the black 
curve its projection on the $(s,w)$-plane, and the red curve is the orbit of the quadrupole.}
\label{fig:05}
\end{figure}

Let us finally consider the trajectory in~\reffig{fig:05} which is far away from both the quadrupole soliton
as well as from the ''turning point'' ($\beta$) by letting $\Gamma=0.5$.
The dynamics is still quasiperiodic and stable (at least up to $t=1500$), but involves multiple frequencies.
Interestingly, the dominant frequency of oscillation with period $T\approx2.6$ can be related to a stable eigenvalue of
the quadrupole soliton $Q$ for $M=200$. In the (stable) eigenvalue spectrum of $Q$ shown in~\reffig{fig:spectrum_M_200}b), the internal mode with $\kappa\approx2.6$ resembles a (modulated) ring with a hump (not shown).
The duration of one period $T$ would then be given by $T=2\pi/\kappa\approx 2.4$, which is what we find when we slightly perturb the quadrupole soliton $Q$ by this mode. Moreover, for $\Gamma=0.1$ (not shown) we also find an oscillation with period $T\approx2.4$. In both case, the propagation dynamics resemble the one shown in~\reffig{fig:05} for $\Gamma=0.5$.
Thus, even though for $\Gamma=0.5$ we are no longer in the region where perturbation analysis of the quadrupole soliton $Q$ holds, we still find qualitatively similar dynamics. We note that in the same system~\refeq{eq:NLS}, quasiperiodic nonlinear solutions (so-called azimuthons) linked to stable internal modes of solitons were reported earlier~\cite{Skupin:oe:16:9118,Maucher:oqel:2009}.

To sum up, we have identified a family of stable quasiperiodic solutions to \refeq{eq:NLS}, starting from $\psi_\Gamma$ given in~\refeq{eq:psi_alpha} and $0<\Gamma<1$. The two limiting solutions are the stable quadrupole solitons $Q$ ($\Gamma=0$) and the homoclinic orbit linked to the unstable radial solitons $R_2$ ($\Gamma=1$). We want to emphasize here that for lower masses, where the quadrupole soliton $Q$ becomes unstable (e.g., $M=85$), we were not able to find stable quasiperiodic solutions by the same construction.

\subsection{Fourier spectrum}\label{fourier}

\begin{figure}
\centerline{\includegraphics[width=\columnwidth]{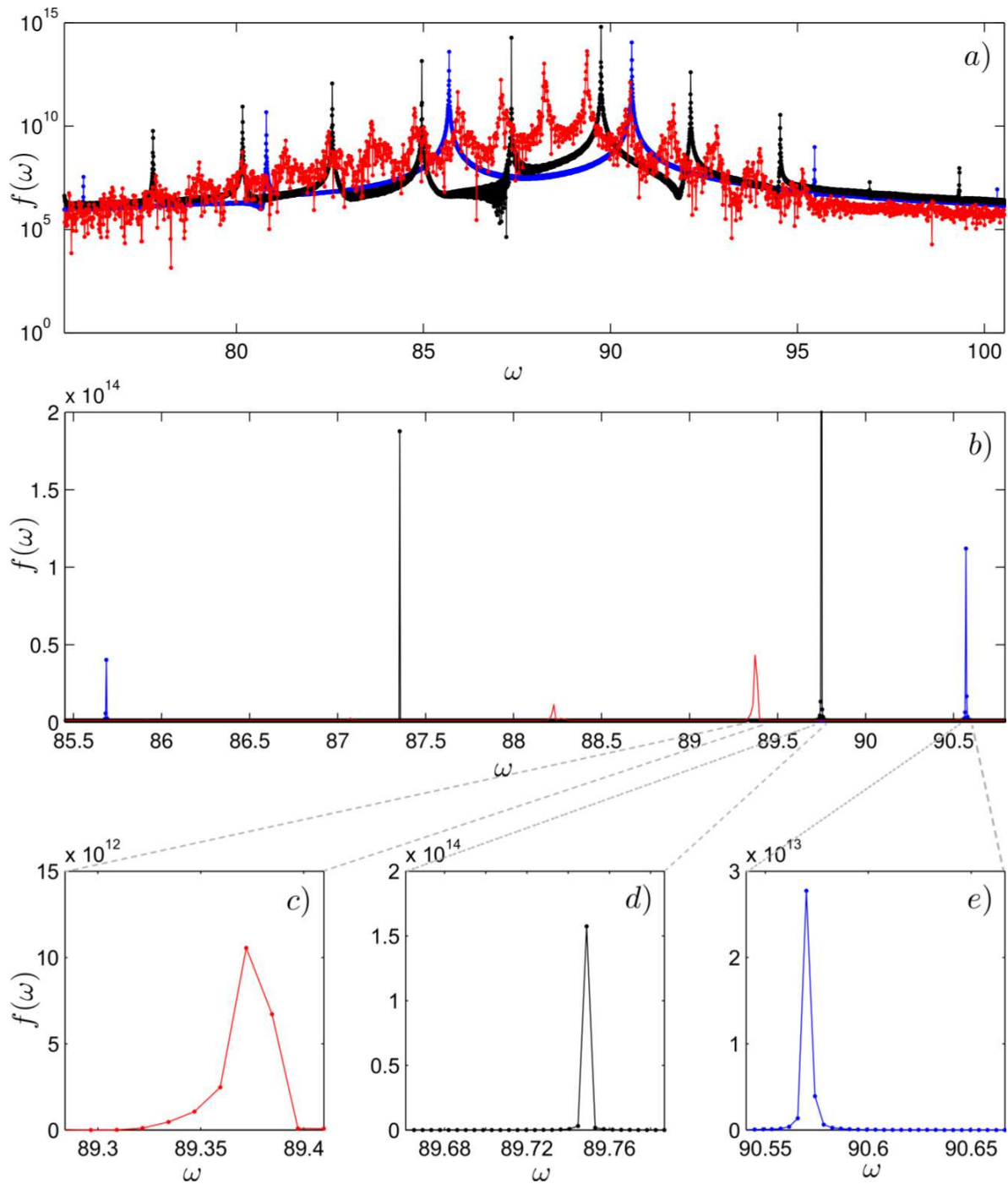}}
\caption{a) Spectrum $f(\omega)=|\mathcal{F}(\psi(\vec{r}=0,t)|^2$ corresponding to the homoclinic orbit $\Gamma=1.00$ (red), and quasiperiodic orbits with $\Gamma=0.99$ (black) and $\Gamma=0.5$ (blue) in logarithmic scale. b) Same information in linear scale. c)--e) show magnifications of single peaks of b).}
\label{fig:spectrum}
\end{figure}

Further insight can be gained by considering the Fourier spectrum of the above trajectories.
Given a trajectory $\psi(\vec{r},t)$ we compute the modulus of the Fourier transform $\mathcal{F}$ of the wavefunction at a fixed point in space (in our case the origin $\mathbf{r}=0$):
\begin{equation}
 f(\omega) = |\mathcal{F}(\psi(\vec{r}=0,t)|^2.
\end{equation}
For a bright soliton solution of the form~\refeq{eq:Phi}, one would expect $f(\omega)$ to comprise of a single sharp peak at $\omega=\lambda$.
On the other hand, in the case of quasiperiodic dynamics with vibration frequency $\Omega$ and propagation constant $\lambda$,
one would expect peaks at $\lambda+m\,\Omega$, where $m$ is integer. This is readily verified for the orbits with $a=0.99$
and $a=0.5$, as can be seen in \reffig{fig:spectrum}, where we see sharp peaks associated with these orbits.
On the other hand, there is no well defined periodicity associated with the homoclinic orbit, since the time spent in the vicinity of
$R_2$ is in principle infinite. In practice, this time is greatly affected by numerical noise and the spectrum appears continuous [see
\reffig{fig:spectrum}a)]. Even if it is possible to associate a dominant frequency $\Omega$ with the homoclinic orbit,
$f(\omega)$ around $\Omega$ is much broader than in the case of quasiperiodic orbits for $\Gamma=0.99$ and $\Gamma=0.5$ [see \reffig{fig:spectrum}b) and c)]\footnote{A limitation
on the spectral resolution for $f(\omega)$ for the homoclinic orbit appears due to the fact that dynamics become unstable around $t=520$. Here, we used
the interval $t=[0:500]$ to compute the spectrum.
Thus, compared to the other two spectra shown in~\reffig{fig:spectrum}, where the propagation was performed until $t=1500$,
the spectral resolution is coarser by a factor of three.}.

Thus, the Fourier spectra yield an additional indication of the qualitatively different nature of the dynamics of Sec.~\ref{sec:homo}
from the quasiperiodic motion of Sec.~\ref{sec:quasi}, providing further support for the conjectured existence of
an underlying homoclinic connection in the former case.

\section{Conclusions} \label{secV}

In previous works, an oscillatory shape-transformation of modes in nonlocal media has been observed~\cite{Bocculiero:PRL:2007,Buccoliero:OE:09}.
In this paper, we approached this phenomenon by means of linear stability analysis and projection techniques borrowed from dynamical systems studies of dissipative PDEs. By studying the linear stability of the quadrupole soliton $Q$ and the second-order radial soliton $R_2$,
we found that the former becomes linearly stable for mass $M \gtrsim 90$,
whereas the latter remains linearly unstable for all masses.
The initial stage of the shape-transformations under consideration, i.e. the emergence of a new state on top of $R_2$,
can be understood in terms of this linear instability, which is
triggered by the unavoidable numerical noise.
However, the most striking feature of the dynamics, i.e. the return to the initial state, is inherently nonlinear, as  it occurs only after the
linear instability saturates.
To study this phenomenon, we introduced a low-dimensional representation of the dynamics, through a projection to dynamically
important states, which were constructed from the radial soliton $R_2$ itself and its unstable/stable eigenmodes.
Projecting the time evolution of the wavefunction $\psi(\vec{r},t)$ (obtained by integrating the NLS) onto these states allows a visualization of oscillatory shape-transformations in terms of trajectories, revealing that shape-transformations
can be interpreted as a homoclinic orbit leaving and re-approaching $R_2$.
Moreover, in the neighborhood of this homoclinic orbit we found quasiperiodic solutions, which for small enough perturbations
resemble the homoclinic connection. This indicates that the homoclinic connection provides a basic recurrence mechanism around
which quasiperiodic dynamics is organized, as is common in lower-dimensional dynamical systems~\cite{book_nonlinear_dynamics}. We were also able to construct and identify a whole family of stable quasiperiodic orbits when the quadrupole soliton $Q$ is stable.

The projection method introduced here allows a compact representation of the dynamics, dual to the commonly used intensity plots.
Moreover, in certain cases it helps to uncover features of the dynamics that are not apparent in snapshots of the intensity evolution.
We expect that similar studies can be carried out for other states exhibiting similar dynamics~\cite{Bocculiero:PRL:2007} and that
our projection method (or similar extensions of the methods of Refs.~\cite{GHCW07,SCD07})
could be applied to a variety of high- and infinite-dimensional conservative systems.

\section*{References}
\bibliographystyle{unsrt}
\bibliography{quasiperiodic}

\begin{thebibliography}{10}

\bibitem{Agrawal:book:2006}
G.~P. Agrawal.
\newblock {\em Nonlinear Fiber Optics}.
\newblock Academic Press, San Diego, third edition, 2001.

\bibitem{Sulem:book:1999}
C.~Sulem and P.-L. Sulem.
\newblock {\em The Nonlinear {Schr{\"o}dinger} Equation: Self-focusing and Wave
  collapse}.
\newblock Springer-Verlag, New York, first edition, 1999.

\bibitem{Litvak:JETP:1966}
A.~G. Litvak.
\newblock Self-focusing of powerful light beams by thermal effects.
\newblock {\em JETP Lett.}, 4:230, 1966.

\bibitem{Litvak:1975}
A.~G. Litvak, V.~A. Mironov, G.~M. Fraiman, and A.~D. Yunakovskii.
\newblock Thermal self-effect of wave beams in a plasma with a nonlocal
  nonlinearity.
\newblock {\em Sov. J. Plasma Phys.}, 1:31--37, 1975.

\bibitem{Davydova:ujp:40:487}
T.~A. Davydova and A.~I. Fishchuk.
\newblock Upper hybrid nonlinear wave structures.
\newblock {\em Ukr.\ J.\ Phys.}, 40:487, 1995.

\bibitem{Wright:85}
E.~M. Wright, W.~J. Firth, and I.~Galbraith.
\newblock Beam propagation in a medium with a diffusive {Kerr-type}
  nonlinearity.
\newblock {\em J. Opt. Soc. Am. B}, 2:383--386, 1985.

\bibitem{Ultanir:OL:04}
E.~A. Ultanir, G.~I. Stegeman, C.~H. Lange, and F.~Lederer.
\newblock Coherent interactions of dissipative spatial solitons.
\newblock {\em Opt. Lett.}, 29:283--285, 2004.

\bibitem{Happer:PRL:1977}
A.~C. Tam and W.~Happer.
\newblock Long-range interactions between cw self-focused laser beams in an
  atomic vapor.
\newblock {\em Phys. Rev. Lett.}, 38:278--282, 1977.

\bibitem{Suter:PRA:1993}
D.~Suter and T.~Blasberg.
\newblock Stabilization of transverse solitary waves by a nonlocal response of
  the nonlinear medium.
\newblock {\em Phys. Rev. A}, 48:4583--4587, 1993.

\bibitem{Goral:PRA:05}
K.~Goral, K.~Rzazewski, and T.~Pfau.
\newblock {Bose-Einstein} condensation with magnetic dipole-dipole forces.
\newblock {\em Phys. Rev. A}, 61:051601(R), 2000.

\bibitem{Tilman:PRL:2005}
A.~Griesmaier, J.~Werner, S.~Hensler, J.~Stuhler, and T.~Pfau.
\newblock {Bose-Einstein} condensation of chromium.
\newblock {\em Phys. Rev. Lett.}, 94:160401, 2005.

\bibitem{Beaufils:PRA:2008}
Q.~Beaufils, R.~Chicireanu, T.~Zanon, B.~Laburthe-Tolra, E.~Mar\'echal,
  L.~Vernac, J.-C. Keller, and O.~Gorceix.
\newblock All-optical production of chromium {Bose-Einstein} condensates.
\newblock {\em Phys. Rev. A}, 77:061601, 2008.

\bibitem{Santos:PRL:2005}
J.~Stuhler, A.~Griesmaier, T.~Koch, M.~Fattori, T.~Pfau, S.~Giovanazzi,
  P.~Pedri, and L.~Santos.
\newblock Observation of dipole-dipole interaction in a degenerate quantum gas.
\newblock {\em Phys. Rev. Lett.}, 95:150406, 2005.

\bibitem{Henkel:PRL:2010}
N.~Henkel, R.~Nath, and T.~Pohl.
\newblock Three-dimensional roton excitations and supersolid formation in
  {Rydberg-excited} {Bose-Einstein} condensates.
\newblock {\em Phys. Rev. Lett.}, 104:195302, 2010.

\bibitem{Maucher:PRL:2011}
F.~Maucher, N.~Henkel, M.~Saffman, W.~Kr\'olikowski, S.~Skupin, and T.~Pohl.
\newblock Rydberg-induced solitons: Three-dimensional self-trapping of matter
  waves.
\newblock {\em Phys. Rev. Lett.}, 106:170401, 2011.

\bibitem{McLaughlin:PhysicaD:95}
D.~W. McLaughlin.
\newblock A paraxial model for optical self-focussing in a nematic liquid
  crystal.
\newblock {\em Physica D}, 88:55, 1995.

\bibitem{Assanto:IEEEQE:03}
G.~Assanto and M.~Peccianti.
\newblock Spatial solitons in nematic liquid crystals.
\newblock {\em Quantum Electronics, IEEE Journal of}, 39:13 -- 21, 2003.

\bibitem{Conti:PRL:2003}
C.~Conti, M.~Peccianti, and G.~Assanto.
\newblock Route to nonlocality and observation of accessible solitons.
\newblock {\em Phys. Rev. Lett.}, 91:073901, 2003.

\bibitem{Peccianti:OL:05}
M.~Peccianti, C.~Conti, and G.~Assanto.
\newblock Interplay between nonlocality and nonlinearity in nematic liquid
  crystals.
\newblock {\em Opt. Lett.}, 30:415--417, 2005.

\bibitem{Briedis:OE:05}
D.~Briedis, D.~Petersen, D.~Edmundson, W.~Krolikowski, and O.~Bang.
\newblock Ring vortex solitons in nonlocal nonlinear media.
\newblock {\em Opt. Express}, 13:435--443, 2005.

\bibitem{Lopez-Aguayo:OL:06}
S.~Lopez-Aguayo, A.~S. Desyatnikov, Y.~S. Kivshar, S.~Skupin, W.~Krolikowski,
  and O.~Bang.
\newblock Stable rotating dipole solitons in nonlocal optical media.
\newblock {\em Opt. Lett.}, 31:1100--1102, 2006.

\bibitem{Maucher:PRA:2012}
F.~Maucher, W.~Krolikowski, and S.~Skupin.
\newblock Stability of solitary waves in random nonlocal nonlinear media.
\newblock {\em Phys. Rev. A}, 85:063803, 2012.

\bibitem{Turitsyn:tmf:85}
S.~K. Turitsyn.
\newblock Spatial dispersion of nonlinearity and stability of multidimensional
  solitons.
\newblock {\em Theor, Mat. Fiz.}, 64:797--801, 1985.

\bibitem{Bang:pre:2002}
O.~Bang, W.~Krolikowski, J.~Wyller, and J.~J. Rasmussen.
\newblock Collapse arrest and soliton stabilization in nonlocal nonlinear
  media.
\newblock {\em Phys. Rev. E}, 66:046619, 2002.

\bibitem{Maucher:nonlinearity:2011}
F.~Maucher, S.~Skupin, and W.~Krolikowski.
\newblock {Collapse in the nonlocal nonlinear Schrodinger equation}.
\newblock {\em {Nonlinearity}}, {24}:{1987}, {2011}.

\bibitem{Bocculiero:PRL:2007}
D.~Buccoliero, A.~S. Desyatnikov, W.~Krolikowski, and Y.~S. Kivshar.
\newblock Laguerre and {Hermite} soliton clusters in nonlocal nonlinear media.
\newblock {\em Phys. Rev. Lett.}, 98:053901, 2007.

\bibitem{Buccoliero:OE:09}
D.~Buccoliero and A.~S. Desyatnikov.
\newblock Quasi-periodic transformations of nonlocal spatial solitons.
\newblock {\em Opt. Express}, 17:9608--9613, 2009.

\bibitem{GHCW07}
J.~F. Gibson, J.~Halcrow, and P.~Cvitanovi{\'c}.
\newblock Visualizing the geometry of state-space in plane {Couette} flow.
\newblock {\em J. Fluid Mech.}, 611:107--130, 2008.

\bibitem{SCD07}
P.~Cvitanovi{\'c}, R.~L. Davidchack, and E.~Siminos.
\newblock On the state space geometry of the {Kuramoto-Sivashinsky} flow in a
  periodic domain.
\newblock {\em SIAM J. Appl. Dyn. Syst.}, 9:1--33, 2010.

\bibitem{Skupin:PRE:2006}
S.~Skupin, O.~Bang, D.~Edmundson, and W.~Krolikowski.
\newblock Stability of two-dimensional spatial solitons in nonlocal nonlinear
  media.
\newblock {\em Phys. Rev. E}, 73:066603, 2006.

\bibitem{Skupin:oe:16:9118}
S.~Skupin, M.~Grech, and W.~Krolikowski.
\newblock {Rotating soliton solutions in nonlocal nonlinear media}.
\newblock {\em {Opt. Express}}, 16:9118--9131, 2008.

\bibitem{Maucher:oqel:2009}
F.~Maucher, D.~Buccoliero, S.~Skupin, M.~Grech, A.S. Desyatnikov, and
  W.~Krolikowski.
\newblock Tracking azimuthons in nonlocal nonlinear media.
\newblock {\em Optical and Quantum Electronics}, 41:337--348, 2009.

\bibitem{book_nonlinear_dynamics}
J.~Guckenheimer and P.~Holmes.
\newblock {\em Nonlinear Oscillations, Dynamical Systems, and Bifurcations of
  Vector Fields}.
\newblock Springer, New York, 1983.

\end{thebibliography}

\appendix

\section*{Appendix}

\section{Rotational invariance of $U(t)$ and $S(t)$} \label{appeA}

Here we prove that the quantities $U(t),\, S(t)$ are rotationally invariant, i.e. they have the same value
if we substitute $\psi(x,y,t)$ with $\Rot(\theta)\psi(x,y,t)=\psi(x\,\cos\theta-y\sin\theta,x\,\sin\theta+y\cos\theta,t)$,
where $\Rot(\theta)$ is an $\mathrm{SO}(2)$ rotation.

The eigenproblem \refeq{eq:internal_modes} for the ring soliton $R_2$ is rotationally symmetric and, as a result,
its internal modes $\ev{1},\,\ev{2}$ transform according to
\begin{equation}\label{eq:evSO2}
  \Rot(\theta) \ev{i} = \sum\limits_{j=1}^2\mathrm{D}_{ji}(\theta)\,\ev{j}\,,
\end{equation}
where $\mathrm{D}(\theta)$ is a two-dimensional matrix-representation of $\mathrm{SO}(2)$. The explicit
representation $\mathrm{D}(\theta)$
depends on the basis $\ev{j}$, but for our purposes it is sufficient to show that we have a
real representation. We begin by noting that the constraints of
orthogonality, $\mathrm{D}^T\mathrm{D}=1$, and unit determinant, $\det(\mathrm{D})=1$, lead to the following
general form
\begin{equation}\label{eq:SO2rep}
  \mathrm{D}(\theta)= \left(
			\begin{array}{cc}
				\alpha(\theta) & \beta(\theta) \\
				-\beta^*(\theta) & \alpha^*(\theta)
			\end{array}
		      \right)
\end{equation}
where the functions $a(\theta),\,\beta(\theta)$ are related through
\begin{equation}\label{eq:SO2unimod}
    \det\left(\mathrm{D}(\theta)\right)= |\alpha(\theta)|^2+|\beta(\theta)|^2=1.
\end{equation}
On the other hand, using $\ev{2}=\Rot(\theta_0)\ev{1}$, where $\theta_0$ is the angle that rotates $\ev{1}$
onto $\ev{2}$, we can express all matrix elements $\mathrm{D}_{ji}=\scalarp{\ev{j}}{\Rot(\theta)\ev{i}}$ in terms of
$\mathrm{D}_{11}$,
\begin{equation}\label{eq:SO2rep2}
  \mathrm{D}(\theta)= \left(
			\begin{array}{cc}
				\alpha(\theta) & \alpha(\theta+\theta_0) \\
				\alpha(\theta-\theta_0) & \alpha(\theta)
			\end{array}
		      \right)
\end{equation}
Comparing with \refeq{eq:SO2rep} we conclude that $\alpha(\theta)=\alpha^*(\theta)$ and thus our representation is real,
and that $\alpha(\theta-\theta_0)=-\alpha(\theta+\theta_0).$\footnote{In our numerical results $\theta_0=\pi/4$
and one can see that our representation is in fact equivalent to
\[
  \mathrm{D}(\theta)= \left(
			\begin{array}{cc}
				\cos(2\theta) & -\sin(2\theta) \\
				\sin(2\theta) & \cos(2\theta)
			\end{array}
		      \right)\,.
\]
}

Using \refeqs{eq:evSO2}{eq:SO2unimod} in definition \refeq{eq:evperp}, along with the relation
$\scalarp{\ev{1}^*}{\ev{1}}=\scalarp{\ev{2}^*}{\ev{2}}$, one can show that
\begin{equation}\label{eq:evperpSO2}
  \Rot(\theta) \evperp{i} = \sum\limits_{j=1}^2\mathrm{D}_{ji}(\theta)\,\evperp{j}\,.
\end{equation}
Then, using \refeqs{eq:SO2rep}{eq:evperpSO2}, it's easy to show that
\begin{eqnarray*}
  \bar{U}^2(t)	& \equiv \left|\scalarp{\evperp{1}}{\Rot(\theta)\psi}\right|^2+\left|\scalarp{\evperp{2}}{\Rot(\theta)\psi}\right|^2\\
		& = \left|\scalarp{\Rot(-\theta)\evperp{1}}{\psi}\right|^2+\left|\scalarp{\Rot(-\theta)\evperp{2}}{\psi}\right|^2\\
		& = \left|\scalarp{\evperp{1}}{\psi}\right|^2+\left|\scalarp{\evperp{2}}{\psi}\right|^2\\
		& = U^2(t)\,.
\end{eqnarray*}
A similar proof holds for $S(t)$.

\end{document}